\documentclass[nocite]{epl}

\def\ave#1{\langle #1\rangle}
\newcommand{\Ord}[1]{{\cal O}\left(#1\right)}

\newcommand{\dodo}{{\rm DODO}}
\newcommand{\dudu}{{\rm DUDU}}
\newcommand{\coe}{{\rm COE}}
\newcommand{\Ort}{{\rm O}}
\newcommand{\Un}{{\rm U}}
\newcommand{\tr}{{\, \rm tr}}
\newcommand{\diag}{{\, \rm diag}}

\title{Random--Matrix Ensembles for Semi--Separable Systems}
\shorttitle{rmt of semi-separable systems\hspace{1.5mm}}
\author{T. Prosen\inst{1} \and T. H. Seligman\inst{2,3} and
H. A. Weidenm\"uller\inst{3}}

\institute{
\inst{1} 
Physics Department, Faculty of Mathematics and Physics, 
University of Ljubljana, Slovenia\\
\inst{2}
Centro de Ciencias Fisicas, U.N.A.M., Cuernavaca,
and Centro Internacional de Ciencias, Cuernavaca, Mexico\\
\inst{3}
Max Planck Institut f\" ur Kernphysik, Heidelberg, Germany
}

\begin{document}

\maketitle

\begin{abstract}
Many models for chaotic systems consist of joining two integrable 
systems with incompatible constants of motion. The quantum
counterparts of such models have a propagator which factorizes into
two integrable parts. Each part can be diagonalized. The two
eigenvector bases are related by an orthogonal (or unitary)
transformation. We construct a random matrix ensemble that mimics this
situation and consists of a product of a diagonal, an orthogonal, 
another diagonal and the transposed orthogonal matrix. The diagonal
phases are chosen at random and the orthogonal matrix from Haar's
measure. We derive asymptotic results (dimension $N\to \infty$) using
Wick contractions. A new approximation for the group integration
yields the next order in $1/N$. We obtain a finite correction to the
circular orthogonal ensemble, important in the long--range part of
spectral correlations.
\end{abstract}

It is usually assumed that the spectral fluctuation properties of
classically chaotic systems coincide with those of the corresponding
canonical ensemble of random--matrix theory (``quantum chaos
conjecture''). There is abundant numerical evidence for the
conjecture. In addition, several approaches have aimed at an
analytical proof for the conjecture, see Section 5.9 of the review
\cite{guhr}. Naturally, these approaches have addressed generic
systems. However, many of the commonly studied chaotic systems have a
very particular form: They can be divided into two parts each of which
is integrable, albeit in different coordinates. Such systems have been
called semi--separable \cite{prosen}. Typical cases are: The quarter
stadium, kicked systems such as the kicked rotor, chaotic Jung
scattering maps for integrable Hamiltonians \cite{jung}, or Lombardi's
approximation for Rydberg molecules \cite{lombardi1}. The spectral
fluctuations of such semi--separable systems have been found to be
essentially consistent with the quantum chaos conjecture. In view of
the special nature of these systems, that finding is somewhat
surprising. In the present letter, we construct two random--matrix
models (one for the unitary and one for the orthogonal case) for
semi--separable systems which we can solve analytically. The models
take account of very specific features of semi--separable systems
which are not reproduced by the circular ensembles of random--matrix
theory. With the help of these models, we show why semi--separable 
systems nearly follow the predictions of standard random--matrix
theory, and we predict quantitatively the deviations that are
typically expected for such systems. In order to perform the ensemble
averages, we develop a novel technique of approximate integration
over the orthogonal or the unitary group in the limit of large matrix
dimension.

Semi--separable systems are computationally relatively simple: One
uses a surface of section that separates the two integrable parts. For
kicked rotors one gets an intertwining of gauge transformation and
Fourier transformation if (time) sections are taken before and after
each kick. For scattering maps one obtains transport with the full and
the free Hamiltonian if the cut is placed in the asymptotic region. In
the case of time--independent systems with two degrees of freedom, one
can often choose a Poincar\'e section such that both parts become
integrable scattering systems when the other part is replaced by a
scattering channel. The quantum Poincar\'e map (QPM) is then the
product of two Jung scattering maps \cite{smilansky,prosen2}. The
latter can be unitarised \cite{prosen}. In particular cases such as
the quantum--defect theory of the Rydberg molecule, the division may
relate to a variable such as angular momentum that has a discrete
quantum number, leading to a unitary QPM \cite{lombardi2}. For
surfaces of section, the relevant ensembles to implement the
quantum chaos conjecture are the circular orthogonal and 
the circular unitary ensemble (COE and CUE).

Typically, the time--evolution operator or the QPM of a
semi--separable time--reversal invariant system consists of a sequence
of four operations: The orthogonal transformation $O$ which takes us
from the basis where one part of the propagator is diagonal to the one
where the other part of the propagator is diagonal, followed by a
diagonal phase matrix $D_1 = \diag\{e^{2i \eta_n}\}$, this by the
inverse transformation matrix $O^T$ and this by another diagonal phase
matrix $D_2=\diag\{e^{2i \xi_n}\}$. For the propagator $T'$ this
yields the $N\times N$ matrix $T' = D_2 O^T D_1 O$. The
time--evolution operator can equivalently be written as
\begin{equation}
\label{eq1}
T = D_2^{-1/2} T' D_2^{1/2} = D_{2}^{1/2} O^T D_1 O D_2^{1/2} \ .
\end{equation}
If time--reversal invariance is broken in either of the two
sub--systems, the orthogonal matrix $O$, and its inverse $O^T$, have 
to be replaced by a unitary matrices $U$, and $U^\dagger=U^{-1}$. 
The ensembles are then generated by assuming that
the phases in $D_1$ and $D_2$ are independent random variables, i.e.,
that they obey Poissonian statistics, and that $O$ and $U$ are members
of ensembles defined by the Haar measure for the orthogonal and
unitary matrices, respectively. We denote the resulting two ensembles
for $T$ by DODO and DUDU, respectively. Note that such and similar
composed ensembles of random matrices have been studied numerically 
in Ref.\cite{zyc}.

We determine the spectral properties of DODO and DUDU by first
computing the spectral form factors $K (m) = \ave{|\tr T^m|^2}$ for
both cases. From $K (m)$ we can calculate any 2--point statistics. For
example, the number variance $\Sigma^2(L;N)$ is given by
\begin{equation}
\Sigma^2(L;N) = \frac{2}{\pi^2} \sum_{m=1}^\infty \frac{1}{m^2} 
\sin^2\left(\frac{m\pi L}{N}\right) K(m) \ .
\label{eq:sigform}
\end{equation}
Here $L$ is the length of the spectrum over which the variance is
calculated. To leading order in $1/N$, the asymptotic result for DODO
is given by
\begin{equation}
\Sigma^2_\dodo(L;N) = \Sigma^2_\coe(L;N) + 
\frac{2}{\pi^2} \sin^2(\pi L/N) + \Ord{1/N} \ ,
\label{eq:asymp}
\end{equation}
and similarly for DUDU.

In order to compute $K (m)$, we must calculate averages of powers of
$T$. For the quadratic term, we have
\begin{equation}
\ave{T_{kl}T_{pq}^{*}}_\dodo = 
\ave{e^{i(\eta_k + \eta_l - \eta_p - \eta_q)}}_\eta
\sum_{a,b} \ave{e^{2i(\xi_a - \xi_b)}}_\xi
\ave{O_{ka}O_{la}O_{pb}O_{qb}}_\Ort \ .
\end{equation}
The averages $\ave{}_\eta$ and $\ave{}_\xi$ are trivial and yield
$\ave{e^{i\eta_k-i\eta_l}}_\eta=\delta_{k,l}$ and correspondingly for
$\ave{}_\xi$. This gives $\ave{T_{kl}T_{pq}^{*}}_\dodo = \Delta(kl,pq)
\sum_{a}\ave{O_{ka}O_{la}O_{pa}O_{qa}}_\Ort$. The factor
$\Delta(kl,pq)$ equals unity if the pairs of indices $kl$ and $pq$
coincide (so that either $k = p$ and $l = q$ or $k = q$ and $l = p$)
and vanishes otherwise. The difficulty lies in calculating the average
$\ave{}_\Ort$ over the orthogonal group $\Ort(N)$. This is done
approximately. We begin with the terms of leading order in $1/N$. We 
assume the matrix elements to be independent variables, and we replace
the group integration by a Gaussian average $\langle \rangle_G$ over
the space of real $N \times N$ matrices,
\begin{equation}
\ave{f(O)}_G = \left(\frac{N}{2\pi}\right)^{N^2/2}
\int \prod_{k,l=1}^N d O_{kl}\; f(O) \exp\left(-\frac{1}{2N}\tr(O^T
O)\right) \ .
\label{eq:gauss}
\end{equation}
For the variances, this yields the correct expression $\ave{O_{i j}
  O_{k l}}_G = \frac{1}{N} \delta_{i k} \delta_{j l}$. In this
approximation, averages of monomials of matrix elements of $O$ can be
calculated by standard Wick contraction rules for Gaussian averages.
Because of the symmetry of the matrix $T$, the only two
non--equivalent non--vanishing quadratic moments of $T$ are
\begin{equation}
\ave{T_{11}T^*_{11}}_\dodo \approx 3/N, \label{eq:U11dodo}
\quad
\ave{T_{12}T^*_{12}}_\dodo \approx 1/N \ . \label{eq:U12dodo}
\end{equation}
Comparing this to the result for COE we find that only the diagonal
term differs to leading order: The COE result is $2/N$. For $K(1)$,
this yields $K(1) = 3$ which leads directly to Eq.~(\ref{eq:asymp}).
Higher moments of $T$ are not affected by the difference because when
compared to non--diagonal contractions, the contribution of diagonal
contractions is suppressed by a factor $1/N$.

In order to obtain the subleading terms in $1/N$, we must improve our
approximate group integration. We note that the Gaussian measure in
Eq.~(\ref{eq:gauss}) yields orthogonality of the matrix $O_{mn}$ only
for quadratic terms, $\ave{(O^T O)_{k l}}_G = \delta_{k,l}$. For
higher polynomials in $O_{mn}$, corrections in the first subleading
order occur. For example,
\begin{equation}
\ave{(O^T O)_{k_1 l_1} (O^T O)_{k_2 l_2}\cdots (O^T O)_{k_m l_m}}_G
= \delta_{k_1 l_1}\delta_{k_2 l_2}\cdots \delta_{k_m l_m} + \Ord{1/N},
\quad m \ge 2 \ .
\label{eq:approxorth}
\end{equation}
We observe that as $N \to \infty$, our approximate integration in
Eq.~(\ref{eq:gauss}) is increasingly restricted to the neighborhood of
the group manifold. To obtain the desired improvement, we introduce an
extra weight factor $w(O)$ which is chosen such that the 
orthogonality relations (\ref{eq:approxorth}) become {\em exact} for
larger values of $m$, or the error terms $\Ord{1/N}$ are replaced by 
terms $\Ord{1/N^2}$, or both. We define the new ``average'' $\ave{}_w$,
\begin{equation}
\ave{f(O)}_w = \ave{f(O)w(O)}_G \ .
\label{eq:wint}
\end{equation}
To be useful, the weight function $w(O)$ has to be of low order in the
matrix elements of $O$ and should have a limited number of free
parameters to adjust. We use the ansatz
\begin{equation}
w(O) = A + B \tr(O^T O) + C \tr(O^T O O^T O) + D [\tr(O^T O)]^2 \ .
\label{eq:4term}
\end{equation}
This expression is of fourth order in $O$ and contains all essentially
different terms up to this order. The use of traces of powers of $O^T
O$ guarantees the symmetry under interchange of indices. Moreover, it
causes $w(O)$ to be invariant under left and/or right multiplications
with an {\em arbitrary orthogonal matrix} $O'$,
\begin{equation}
\label{eq:5term}
w(O) = w(O O') = w(O' O),\quad O' \in \Ort(N) \ .
\end{equation}
We note that $O$ is not a member of the orthogonal group. The
invariance property~(\ref{eq:5term}) is desirable because it
guarantees that the measure on the group manifold remains unchanged
except for a constant, and that areas close to this manifold are not
affected drastically because they will be mapped onto other areas near
the manifold. We determine the free parameters $A,B,C,D$ by imposing
the correct orthogonality properties for terms in $O^T O$ of order 0
(normalization), 1, and 2, 
\begin{eqnarray}
\ave{1}_w &=& 1 \ , \label{eq:ord0} \\
\ave{(O^T O)_{ij}}_w &=& \delta_{i,j} \ , \label{eq:ord1}\\
\ave{(O^T O)_{i,j} (O^T O)_{k,l}}_w &=& \delta_{i,j}\delta_{k,l} \ .
\label{eq:ord2}
\end{eqnarray}
Using the approximate integration, we find for the left--hand side of
Eq.~(\ref{eq:ord2}) the form $f_3(A,B,C,D,N) \delta_{i,j}\delta_{k,l}
+ f_4(A,B,C,D,N) (\delta_{i,k}\delta_{j,l} + \delta_{i,l}\delta_{j,k})$.
This yields $f_3(A,B,C,D,N) = 1$ and $f_4(A,B,C,D,N) = 0$. Together
with Eqs.~(\ref{eq:ord0},\ref{eq:ord1}) one finds four linear
equations for the four unknowns $A,B,C,D$. These were solved using
Mathematica 4. We obtain
\footnote{The routines for averaging of
symbolic sums of monomials of matrix elements over
orthogonal and unitary groups using the approach described here
can be obtained from one of the authors by e-mail request on
{\tt prosen@fiz.uni-lj.si}.}
\begin{equation}
A = 1 - \frac{N^2}{4} \ ,\quad 
B = \frac{N}{2} \ ,\quad
C =-\frac{N^3}{4(N^2 + N-2)} \ ,\quad
D = \frac{N^2}{4(N^2 + N-2)} \ .\quad
\label{eq:abcd}
\end{equation}
The negative signs in $A$ and $C$ imply that $w(O)$ is not positive
definite and, thus, is not a proper measure. Yet the constant on the
group manifold is positive, namely $1+N^2/4$. Therefore, on the part
of the integration area near this manifold where the Gaussian factor
is not small, the weight $w(O)$ is also positive. This fact reflects
the invariance condition (\ref{eq:5term}). 

In addition, we get the bonus that with the new integration (\ref{eq:wint}) 
replacing the Gaussian integration (\ref{eq:gauss}) the correction terms on 
the right--hand side of Eq.~(\ref{eq:approxorth}) for $m\ge 3$ are 
$\Ord{\frac{1}{N^2}}$. We have checked
this statement explicitly for $m=3,4,5$, and we conjecture that it
holds for {\em arbitrary} $m \ge 3$. We have also checked that averages 
$\ave{}_w$ over monomials containing the components $O_{i\alpha}$ of a single 
vector (fixed $\alpha$) up to $10$th power are correct to subleading order
${\cal O}(\frac{1}{N^2})$. We also applied our
procedure up to products of three indentities $(O^T O)_{i j}$ in the postulated relations
(\ref{eq:ord0}) - (\ref{eq:ord2}) with a unique solution for 6th order 
polynomial ansatz for $w(O)$.
For $m \ge 4$, no improvement over the 4th order case (\ref{eq:4term},\ref{eq:abcd})
was found in the approximate orthogonality relations Eq.~(\ref{eq:approxorth}) 
(again within ${\cal O}(\frac{1}{N^2})$) for terms of power higher than 6; 
only the latter became exact.

We calculate the form factor for DODO by averaging over all the
Poissonian phases, and by replacing the group integral by $\ave{}_w$.
For $m=1,2$, this yields
\begin{eqnarray}
K_\dodo (1) &=& \sum_{k,p} \ave{O^4_{kp}}_w = \frac{3N}{N+2} \ , 
\quad  \quad \label{eq:K1dodo}\\
K_\dodo (2) &=& 4\!\!\!\sum_{k_1,k_2,p_1,p_2}\!\!\!\ave{O^2_{k_1 p_1}
O^2_{k_2 p_1} O^2_{k_2 p_2} O^2_{k_1 p_2}}_w
- 4\!\!\sum_{k,p_1,p_2}\!\ave{O^4_{k p_1} O^4_{k p_2}}_w +
\sum_{k,p} \ave{O^8_{k p}}_w \\
&=& 4 - \frac{4}{N} + \Ord{N^{-2}} \ .\quad \label{eq:K2dodo}
\end{eqnarray}
The first moment is exact as the averages involved are of the type to
which the weight factor was adjusted. For the second moment we stipulate 
that $\ave{}_w$ correctly reproduces to subleading order, not only the 
averaged expressions mentioned above, but also the first term on the r.h.s. 
of Eq.~(\ref{eq:K2dodo}).

We determine the number variance from Eq.~(\ref{eq:sigform}),
terminating the sum after the second moment. We expect the result to
give a good account of the long--range correlations. We can calculate
the COE result for finite $N$ in the same approximation either in the
conventional way or using similar techniques, see below. For the
difference we obtain
\begin{equation}
\Sigma^2_\dodo(L;N)-\Sigma^2_\coe(L;N) \approx 
\frac{2}{\pi^2}\left\{\left(1-\frac{4}{N}\right) \sin^2 \left(
\frac{\pi L}{N} \right) + \frac{1}{N}\sin^2\left( \frac{2\pi L}{N}
\right) \right\} \ .
\label{eq:appr}
\end{equation}
To lowest order in $1/N$, this formula agrees with Eq.~(\ref{eq:asymp})
and is exact, because only the first moment of $T$ has a correction of
order zero in $1/N$. We expect that the terms of next order in $1/N$
yield a good representation of the long--range behavior in $L$ of the
number variance. Finite--size effects for the short--range behavior are
difficult to predict, except that we expect them to be small on
general grounds, at least as long as the range $L$ is short compared
to the matrix dimension $N$.

\begin{figure}
\onefigure{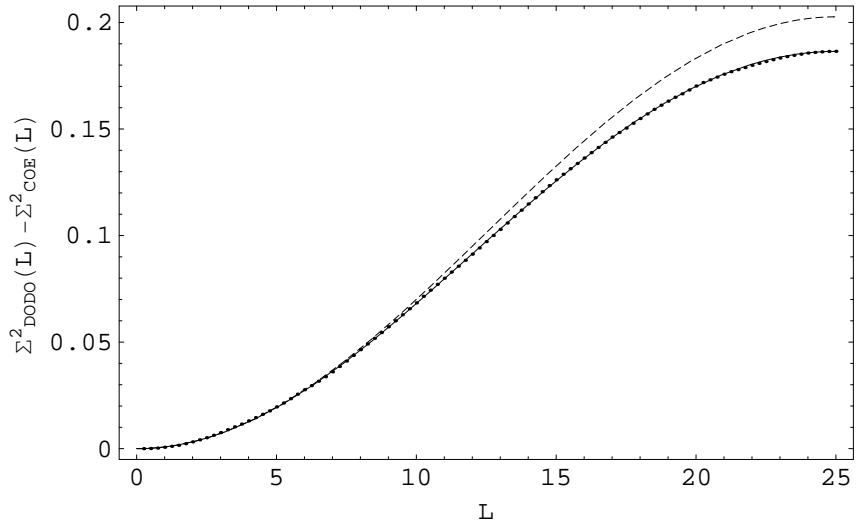}
\caption{Deviation of the number variance of DODO$(N=50)$ from 
COE$(N=50)$ as determined by numerical Monte Carlo simulation (points),
compared to the theoretical prediction with $1/N$
correction~(\ref{eq:appr}) (solid curve) and without $1/N$ correction
($N=\infty$) (dashed curve).}
\label{fig:sigma}
\end{figure}

\begin{figure}
\onefigure{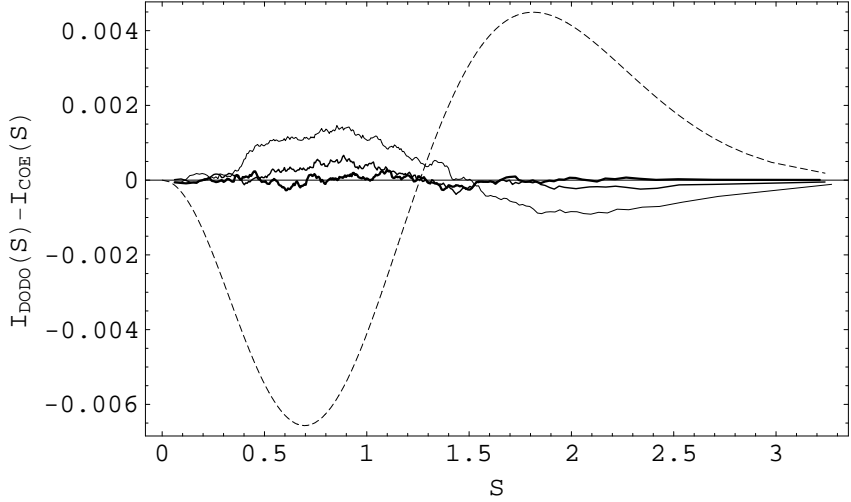}
\caption{Deviation of the integrated level spacing distribution of
DODO$(N)$ from COE$(N)$, $I_{\dodo(N)}(S)-I_{\coe(N)}(S)$, as determined
by numerical simulations for $N=25,50,100$ (thin, medium, thick curve,
respectively). The deviation of the Wigner surmise from COE($\infty$) is
shown by the dashed curve.}
\label{fig:lsd}
\end{figure}

We used numerical calculations to test our results and to look for
finite--size corrections at short distances in the spectrum.
Fig.~\ref{fig:sigma} shows the difference of the number variances for
DODO and COE for three cases: In the limit $N \to \infty $ and for
$N=50$ (in both cases, DODO results were obtained from
Eq.~(\ref{eq:appr}), and the numerical result for $N=50$. We find
excellent agreement (even at short distances) confirming our intuitive
argument. Since the number variance is at short distances not very
sensitive to the changes considered in this paper, we show in
Fig.~\ref{fig:lsd} the difference between the integrated
nearest--neighbour spacing distributions $I(S)=\int_0^S ds P(s)$ for
DODO and COE for $N=25,\; 50,$ and $100$. For purposes of reference,
we have also included a comparison with the Wigner surmise. This shows
how small the correction really is, which for $N=100$ disappears in
the statistical noise of $2\times 10^6$ spacings. We mentioned that the
ensemble we use has been studied numerically before \cite{zyc}. The
authors only looked at fairly short spectral ranges. This is why they
failed to observe the zero order deviation and the $N$--dependence.

For the unitary case, the approximate group integration can be
developed in full analogy to the orthogonal case. For the weight
function $w_c(U)$ (where $U$ is a complex $N$-dimensional matrix) we
use the ansatz $w_c(U) = A_c + B_c \tr(U^\dagger U) + C_c
\tr(U^\dagger U U^\dagger U) + D_c [\tr(U^\dagger U)]^2$ and find
\begin{equation}
A_c = 1 - \frac{N^2}{2} \ ,\quad 
B_c = N \ ,\quad
C_c =-\frac{N^3}{2(N^2-1)} \ ,\quad
D_c = \frac{N^2}{2(N^2-1)} \ .
\end{equation}
Again, the integral of any monomial $\ave{f(U)}_\Un$ is aproximated
up to the first {\em two} (leading and subleading) orders in $1/N$ by
applying complex pairwise contraction rules to the polynomial
$f(U)w_c(U)$. The results for the first two moments for DUDU are
\begin{equation}
K_\dudu(1) = \frac{2N}{N+1} \ ,\quad K_\dudu(2) = 2 - \frac{4}{N}
+ \Ord{\frac{1}{N^2}} \ .
\end{equation}
We have similarly calculated the leading and subleading terms for the
COE and the CUE. The relevant matrices are represented as $U^T U$ and
$U$, respectively. The results coincide with the leading terms of the
$1/N$ expansion of the exact expressions.

Fig.~1 shows that the DODO spectrum is less stiff than that of the
COE. A corresponding statement applies to DUDU versus CUE. The
differences are caused by the Poissonian statistics of the phases
$\eta_k$ and $\xi_a$. To see this, we replace either of the two
diagonal matrices $D_1$ and $D_2$ in Eq.~(\ref{eq1}) by another
diagonal matrix containing the eigenvalues of the COE (or of the
CUE). It is straightforward to show that the resulting ensemble is
identical to the COE (the CUE, respectively). In the opposite limit,
we keep $D_2$ (Poissonian) and replace $D_1$ by $\diag\{\exp(i\alpha n)\}$ 
(picket fence model). We denote the associated first moment by
$K^{p.f.}_\dodo(1)$. Averaging over $\xi_n$, summing over the $n$'s,
and calculating the averages of the $O's$ with the help of the
technique described above, we find
\begin{equation}
\label{eq2}
K^{p.f.}_\dodo(1) = \sum_{m=1}^N\sum_{n=1}^N \exp(i\alpha (m-n))
\sum_{l=1}^N \ave{O_{lm}^2 O_{ln}^2}_\Ort 
= \frac{2N}{N+2} + \frac{\sin^2(N\alpha/2)}{(N+2)\sin^2(\alpha/2)} \ .
\end{equation}
As we deal with a fourth--order monomial, this is an exact result.
Eq.~(\ref{eq2}) yields an interpolation for the long--range level
statistics from the Poissonian value $K(1) = N$ (this applies in the
limit of a degenerate spectrum where $\alpha = 0$) to the COE value
$K_\coe(1) = 2N/(N+2)$ (with error $\Ord{1/N}$) when $\alpha$ becomes
bigger than $2\pi/N$. This result suggests that DODO or DUDU provide
a worst--case scenario for possible deviations from the canonical
random--matrix predictions of COE or CUE.

In summary, we have introduced random--matrix ensembles for
semi--separable systems. These allow us to understand why
semi--separable systems follow the quantum chaos conjecture very
closely, and to predict quantitatively the deviations. On scales $L$
which are of the order of 10\% of the spectral range $N$ or less, the
spectral fluctuations for each ensemble are very similar to those of
the corresponding circular ensemble. Beyond this range, the deviations
reach values of up to 0.2 or so. On a relative scale, these deviations
are small (we recall the logarithmic increase of $\Sigma^2$). On an
absolute scale, they are not. Furthermore, they show a
well--understood dependence on $N$.

In addition, we developed a promising novel way to perform the group
integrals approximately in the limit of large matrix dimension $N$.
Without excessive effort, this method improves the traditional Wick
pair contraction technique and allows for the calculation of
corrections in subleading order. It would be of interest to attain a
deeper understanding of this procedure from a group--theoretical point
of view.

\acknowledgments
We wish to thank F. Leyvraz for useful discussions. This work was 
supported by the ministry of Science of Technology of Slovenia and 
by CONACyT grant 25192-E and by DGAPA (UNAM) grant IN-102597. T.~Prosen 
wishes to express his thanks to the Max Planck Institut f\"ur Kernphysik, 
Heidelberg and Centro Internacional de Ciencias, Cuernavaca for 
their kind hospitality, and T.~H.~Seligman wishes to thank the 
Humboldt Foundation for its support.

\end{document}